\documentclass[preprintnumbers,amsmath,amssymb]{revtex4-1}

\usepackage{epsfig}
\usepackage{graphicx}
\usepackage{subfigure}
\usepackage{float}
\usepackage{multirow}
\usepackage{amssymb}
\usepackage{amsmath}
\usepackage{picinpar}

\begin{document}

\newcommand{\figwidth}{0.9\columnwidth}
\newcommand{\subfigwidth}{0.7\columnwidth}

\newcommand{\smin}{{\mbox{\scriptsize{min}}}}
\newcommand{\smax}{{\mbox{\scriptsize{max}}}}
\newcommand{\stot}{{\mbox{\scriptsize{tot}}}}
\newcommand{\tot}{{\mbox{\scriptsize{tot}}}}

\newcommand{\ms[2]}{{\mbox{\scriptsize{#2}}}}
\newcommand{\im}{i}
\newcommand{\jm}{j}

\newcommand{\mA}{\langle m_A\rangle}
\newcommand{\mB}{{\langle m_B\rangle}}
\newcommand{\pmB}{{\langle pm_B\rangle}}
\newcommand{\m}{{\langle m_\tot\rangle}}

\title{Monte Carlo study of an anisotropic Ising multilayer with antiferromagnetic interlayer couplings}

\author{I. J. L. Diaz}\email{ianlopezdiaz@gmail.com}
\affiliation{Departamento de F\'{i}sica,
Universidade Federal de Santa Catarina,
88040-900, Florian\'{o}polis, SC, Brazil}
\author{N. S. Branco}\email{nsbranco@gmail.com}
\affiliation{Departamento de F\'{i}sica,
Universidade Federal de Santa Catarina,
88040-900, Florian\'{o}polis, SC, Brazil}

\date{\today}

\begin{abstract}
We present a Monte Carlo study of the magnetic properties of an Ising multilayer ferrimagnet.
The system consists of two kinds of non-equivalent planes, one of which is site-diluted.
All intralayer couplings are ferromagnetic.
The different kinds of planes are stacked alternately and the interlayer couplings are antiferromagnetic.
We perform the simulations using the Wolff algorithm and employ multiple histogram reweighting
and finite-size scaling methods to analyze the data with special emphasis on the study of compensation phenomena.
Compensation and critical temperatures of the system are obtained as functions of the Hamiltonian parameters
and we present a detailed discussion about the contribution of each parameter
to the presence or absence of the compensation effect.
A comparison is presented between our results and those reported in the literature
for the same model using the pair approximation.
We also compare our results with those obtained through both
the pair approximation and Monte Carlo simulations for the bilayer system.
\end{abstract}


\maketitle

\section{Introduction}\label{introduction}


The study of ferrimagnetic materials
has attracted considerable attention in the last few decades,
especially since a number of phenomena associated with these materials
present a great potential for technological applications
\cite{connell1982magneto, grumberg1986layered, camley1989theory, phan2007review}.
In such systems the different temperature behavior of the
sublattice magnetizations may cause the appearance of compensation points,
i. e., temperatures below the critical point for which the total magnetization is zero
while the individual sublattices remain magnetically ordered \cite{cullity2011introduction}.
Although unrelated to critical phenomena, at the compensation point 
there are physical properties, such as the magnetic coercivity of the system,
that exhibit a singular behavior \cite{connell1982magneto, shieh1986magneto, ostorero1994dy}.
The fact that the compensation point of some ferrimagnets occurs near room temperature
makes them particularly important for applications in magneto-optical drives \cite{connell1982magneto}.

Initially, compensation effects were theoretically studied in bipartite lattices with different spin
magnitudes in each sublattice \cite{boechat2002renormalization, godoy2004mixed}.
However, this is not the only possible geometry which may lead to compensation phenomena.
In particular, layered ferrimagnets have been extensively studied in the recent past \cite{balcerzak2014ferrimagnetism, szalowski2014normal, diaz2016monte}.
In the latter case, systems are composed of stacked planes with different magnetic properties
and the realization of antiferromagnetic couplings between adjacent layers
has important technological applications such as in
magneto-optical recordings \cite{connell1982magneto},
spintronics \cite{grumberg1986layered},
the giant magnetorresistance \cite{camley1989theory},
and the magnetocaloric effect \cite{phan2007review}.
In addition, the study of the magnetic properties of these systems
is of great theoretical interest since it can provide insight into the
crossover between the characteristic behavior of two- and three-dimensional magnets.

In recent experimental works, we find examples of
realization and study of such
bilayer \cite{stier2011carrier},
trilayer \cite{smits2004antiferromagnetic, leiner2010observation},
and multilayer \cite{kepa2001antiferromagnetic,
chern2001antiparallel,
sankowski2005interlayer,
chung2011investigation,
samburskaya2013magnetization} systems.
From the theoretical stance,
a bilayer system with Ising spins and no dilution
has been studied via
transfer matrix (TM) \cite{lipowski1993layered, lipowski1998critical},
renormalization group (RG) \cite{hansen1993two, li2001critical, mirza2003phenomenological},
mean-field approximation (MFA) \cite{hansen1993two},
and Monte Carlo (MC) simulations \cite{ferrenberg1991monte, hansen1993two}.
A similar system with both Ising and Heisenberg spins
has been studied in the pair approximation (PA)
both without dilution \cite{szalowski2013influence} and with dilution \cite{balcerzak2014ferrimagnetism}.
There is also a recent work considering an Ising bilayer with site dilution
in an MC approach \cite{diaz2016monte}.
For the multilayer system,
the PA has also been applied to the model with Ising and Heisenberg spins,
both with no dilution \cite{szalowski2012critical}
and with dilution \cite{szalowski2014normal}.

If all layers have the same spin (e. g., spin-$1/2$) and we have an even number of layers,
it is necessary that different layers have different number of spins in them
for the existence of a compensation effect, as discussed in Refs. \onlinecite{balcerzak2014ferrimagnetism, szalowski2014normal, diaz2016monte}.
It is also necessary that the layers have asymmetric intralayer exchange integrals
and that the layers with stronger exchange integrals have less atoms than
their weak-exchange counterparts. That is easy to achieve with site dilution.
However, to the best of our knowledge, no numerical simulation methods have yet been applied 
to multilayer systems with site dilution.

With that in mind, in this work we present a Monte Carlo study of the magnetic properties of a spin-$1/2$ Ising system composed of two
kinds of non-equivalent planes, \textbf{A} and \textbf{B}, stacked alternately.
All intralayer interactions are ferromagnetic while the interlayer interactions are antiferromagnetic.
We also consider the presence of site dilution in one of the kinds of planes.
The simulations are performed with the Wolff algorithm \cite{artigo:wolff}
and with the aid of a reweighting multiple histogram technique \cite{artigo:ferrenberg:histograma1, artigo:ferrenberg:histograma2}.
The model is presented in Sec. \ref{model},
the simulation and data analysis methods are discussed in Sec. \ref{monte_carlo},
and the results are presented and discussed in Sec. \ref{results}.

\section{Model and observables}\label{model}

\begin{figure}[h]
\begin{center}
\includegraphics[width=\textwidth]{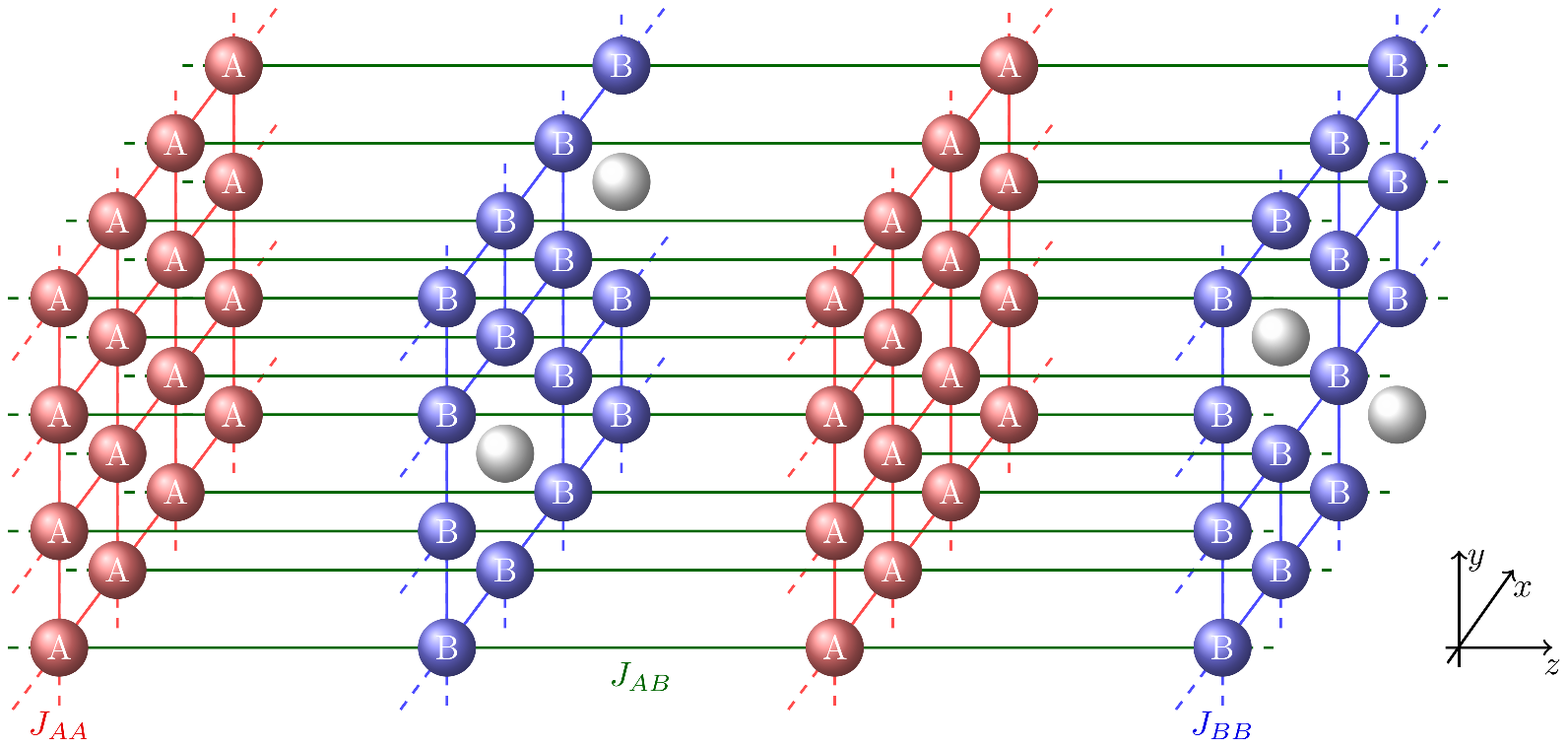}
\caption{
\label{fig:multilayer}
A schematic representation of the multilayer containing an infinite number of layers of types \textbf{A} and \textbf{B} arranged alternately.
The intraplanar exchange integrals for two neighboring atoms belonging to the same layer
are $J_{AA}>0$ (layer \textbf{A}) and $J_{BB}>0$ (layer \textbf{B}).
The interplanar exchange integral is $J_{AB}<0$.}
\end{center}
\end{figure}

The multilayer system we study consists of a simple cubic crystalline lattice
such that non-equivalent monolayers (\textbf{A} and \textbf{B}) are stacked alternately (see Fig. \ref{fig:multilayer}).
The \textbf{A} planes are composed exclusively of \textbf{A}-type atoms while the
\textbf{B} planes have \textbf{B}-type atoms as well as non-magnetic impurities.
The Hamiltonian describing our system is of the Ising type with spin $1/2$ and can be written as
\begin{align}\label{eq:hamiltonian}
-\beta\mathcal{H} =
\sum_{\langle i\in A,j\in A\rangle}K_{AA}s_i s_j
+\sum_{\langle i\in A,j\in B\rangle}K_{AB}s_i s_j\epsilon_j
+\sum_{\langle i\in B,j\in B\rangle}K_{BB}s_i s_j\epsilon_i\epsilon_j,
\end{align}
where the sums run over nearest neighbors, $\beta\equiv (k_BT)^{-1}$, $T$ is the temperature, $k_B$ is the Boltzmann constant,
the spin variables $s_i$ assume the values $\pm 1$,
the occupation variables $\epsilon_i$ are uncorrelated quenched random variables which take on the values
$\epsilon_i=1$ with probability $p$ (spin concentration) or $\epsilon_i=0$ with probability $1-p$ (spin dilution).
The couplings are $K_{AA}>0$ for an \textbf{AA} pair,
$K_{BB}>0$ for a \textbf{BB} pair,
and $K_{AB}<0$ for an \textbf{AB} pair.
The corresponding exchange integrals (see Fig. \ref{fig:multilayer})
are given by $J_{\gamma\delta}=\beta^{-1}K_{\gamma\delta}$, where $\gamma = A, B$ and $\delta = A, B$.

For the purpose of the numerical analysis to be discussed in Sec. \ref{monte_carlo},
we define some observables to be measured in our simulation.
Namely the dimensionless extensive energy is given by $E\equiv \mathcal{H}/J_{BB}$,
and the magnetizations of \textbf{A}-type atoms and \textbf{B}-type atoms are, respectively
\begin{align}
	\label{eq:mA}
	m_A=\frac{1}{N_A}\sum_{i\in A}s_i,
	\\
	\label{eq:mB}
	m_B=\frac{1}{N_B}\sum_{j\in B}s_j\epsilon_j,
\end{align}
where
$N_A=L^3/2$ is the total numbers of \textbf{A}-type atoms in the system and $N_B=pL^3/2$ is the number of \textbf{B}-type atoms.
The total magnetization of the system is
\begin{equation}\label{eq:m}
	m_\tot
	=\frac{1}{2}\left(m_A+pm_B\right).
\end{equation}

We also define the magnetic susceptibilities
\begin{align}
	\label{eq:sus}
	\chi_\gamma
	=N_\gamma K\overline{\left(\langle m_\gamma^2\rangle-\langle |m_\gamma|\rangle^2\right)},
\end{align}
where $\langle\cdots\rangle$ is the thermal average for a single disorder configuration
whereas the over-line denotes the subsequent average over disorder configurations,
$K=\beta J_{BB}$ is the inverse dimensionless temperature, and $\gamma=A, B, \tot$.
The total number of atoms in the system is $N_{\tot}=N_A+N_B$.

\section{Monte Carlo methods}\label{monte_carlo}

\subsection{Simulational details}\label{simulations}

We employed the Wolf single-cluster algorithm \cite{artigo:wolff}
for the MC analysis of Hamiltonian \eqref{eq:hamiltonian}
on cubic lattices of size $L^3$ with periodic boundary conditions.
We performed simulations for linear sizes $L$ from 10 to 100
and for a range of values of the Hamiltonian parameters:
$0.0<p\leq 1.0$, $0.0<J_{AA}/J_{BB}\leq 1.0$, and $-1.0\leq J_{AB}/J_{BB}<0.0$.
All random numbers were generated using the Mersenne Twister pseudo-random number generator \cite{artigo:mersenne-twister}.

For each set of values chosen for the parameters above,
we performed simulations in a range of temperatures close to either the critical point or the compensation point.
To determine $T_{comp}$ we typically divided the range in 5 to 10 equally spaced temperatures
whereas for $T_c$ we used from 8 to 17 equally spaced temperatures.
In a single simulation for the largest systems considered, i. e., $L=100$,
we performed up to $2\times 10^5$ steps (Wolff single-cluster updates)
and discarded up to $7\times 10^4$ steps to account for equilibration.
The total number of steps after equilibration was always at least
2000 times the relevant integrated autocorrelation time \cite{livro:barkema}.

To analyze the data generated in the simulations for a particular range of temperatures,
we use the multiple-histogram method \cite{artigo:ferrenberg:histograma1,artigo:ferrenberg:histograma2}
to compute the observables defined in the previous section at any temperature inside this range.
The thermal error associated with those observables is estimated via the blocking method \cite{livro:barkema},
in which we divide the data from each simulation in blocks and repeat the multiple-histogram procedure
for each block. The errors are the standard deviation of the values obtained
for a given observable for different blocks.

The process is repeated for $N_s$ samples of quenched disorder to obtain the final estimate of our observable.
We chose $10\leq N_s\leq 50$, such that the error due to disorder was approximately
the same as the thermal error obtained for each disorder configuration.
Finally, we sum both thermal and disorder errors for an estimate of the total error.

\subsection{Determining $T_{c}$}\label{sec:tc}

In this work we wish to determine the critical point accurately.
This is accomplished by means of a finite-size scaling analysis \cite{livro:julia},
in which we examine the size dependency of certain observables measured for finite systems
of several sizes and extrapolate these results to the thermodynamic limit, i. e., $L\rightarrow\infty$.
In this approach, the singular part of the free energy density for a system of size $L$,
near the critical point, is given by the scaling form
\begin{equation}\label{eq:RGf2}
	\bar f_{\ms{sing}}(t,h,L)
	\sim L^{-(2-\alpha)/\nu}f^0(tL^{1/\nu}, hL^{(\gamma+\beta)/\nu})
\end{equation}
where $t$ is the reduced temperature, $t=(T-T_c)/T_c$, $T_c$ is the critical temperature of the infinite system,
and $h$ is the external magnetic field given in units of $k_BT$.
The critical exponents $\alpha$, $\beta$, $\gamma$ and $\nu$ are the traditional ones
associated, respectively, with
the specific heat, magnetization, magnetic susceptibility, and correlation length.
Various thermodynamic properties can be determined from Eq. \eqref{eq:RGf2}
and other observables have their corresponding scaling forms, e. g.,
for the magnetic susceptibility at null magnetic field we have
\begin{equation}
	\chi_\tot = L^{\gamma/\nu}
	\mathcal{X}(x_t),
	\label{FSS_chi}
\end{equation}
where $x_t\equiv tL^{1/\nu}$ is the temperature scaling variable.

Eq. \eqref{FSS_chi} provides a powerful method to determine the critical point.
It is clear from this scaling law that $\chi_\tot$ diverges at the critical point
only in the thermodynamic limit, whereas for a finite system size $L$,
$\chi_\tot$ does not diverge but has a maximum at a pseudo-critical temperature $T_c(L)$,
which asymptotically approaches the real $T_c$ as $L$ increases.
The maximum occurs when
\begin{equation}
	\left.\frac{d\mathcal{X}(x_t)}{dx_t}\right|_{T=T_c(L)}=0,
\end{equation}
which yields the following relation
\begin{equation}
	\label{eq:FSS:tc}
	T_c(L) = T_c+AL^{-1/\nu},
\end{equation}
where $A$ is a constant, $T_c$ is the critical temperature and $\nu$ is the
critical exponent associated with the correlation length.

The finite-size scaling method is applicable to different quantities,
such as specific heat and other thermodynamic derivatives \cite{artigo:landau}.
We expect the results obtained from the scaling behavior of these other quantities
to be consistent, as we were able to verify in preliminary simulations.
In this study, however, we focused only on the peak temperatures of the
magnetic susceptibilities, defined in Eq. (\ref{eq:sus}),
for these peak temperatures occurred fairly close to one another and were the
sharpest peaks from all the quantities initially considered.

\begin{figure}[h]
\begin{center}
\includegraphics[width=\figwidth]{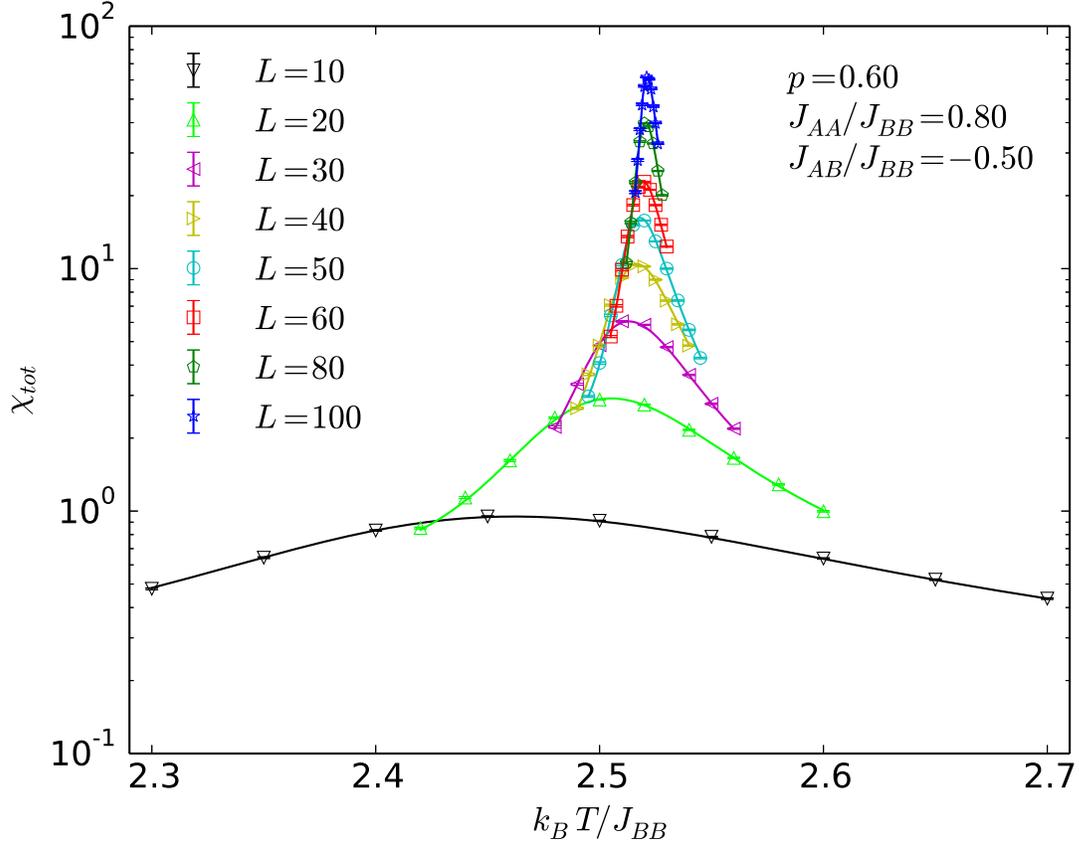}
\caption{
\label{fig:mhist:sus}
Semilog plot of the magnetic susceptibility $\chi_{tot}$ as a function of the dimensionless temperature $k_BT/J_{BB}$
for $J_{AA}/J_{BB}=0.80$, $J_{AB}/J_{BB}=-0.50$, $p=0.60$, and linear lattice sizes $L$ ranging from 10 to 100.
The symbols correspond to simulation data and the solid lines were obtained using the multiple histogram method.
Where the error bars are not visible, they are smaller than the symbols.
}
\end{center}
\end{figure}

To determine the pseudo-critical temperatures $T_c(L)$,
we perform simulations close to the peaks of the susceptibilities
and use the multiple-histogram method to obtain $\chi_A$, $\chi_B$, and $\chi_\tot$
as continuous functions of $T$,
as shown in Fig. \ref{fig:mhist:sus} for the total magnetic susceptibility
of a system with $p=0.60$, $J_{AA}/J_{BB}=0.80$, $J_{AB}/J_{BB}=-0.50$, and $L$ from 10 to 100.
The location of the peak temperatures is
automated using the Broyden-Fletcher-Goldfarb-Shanno (BFGS) method \cite{artigo:BFGS}
for each of the $N_s$ disorder configurations and the errors are estimated as discussed in Sec. \ref{simulations}.

Finally, to obtain $T_c$, we perform least-square fits with Eq. \eqref{eq:FSS:tc}
using the estimates for $T_c(L)$ as input.
This equation has three free parameters to be adjusted in the fitting process and requires great statistical resolution
in order to produce stable and reliable estimates of the parameters.
Since the precise estimation of all of these parameters would lead to a drastic increase in computational work,
and since for the present work we are not interested in a precise value for the exponent $\nu$,
we employ the same procedure presented in Ref. \onlinecite{diaz2016monte},
in which we set a fixed value for the exponent $\nu$ and perform fits with two free parameters, instead of three.
These fits are made, for a fixed value of $\nu$, for system sizes not smaller than $L_{\smin}$
and the value of $L_\smin$ that gives the best fit is located, i. e.,
the one that minimizes the reduced weighted sum of errors, $\chi^2/n_{DOF}$,
where $n_{DOF}$ is the number of degrees of freedom.
Next, we keep changing the values of $\nu$ and $L_\smin$ iteratively until we locate the set of values
that globally minimizes $\chi^2/n_{DOF}$
and use these values to determine our best estimate of $T_c$.
This procedure effectively linearizes the fit, although it does not allow for
an individual error estimate for the exponent $\nu$.

\begin{figure}[h]
\begin{center}
\includegraphics[width=\figwidth]{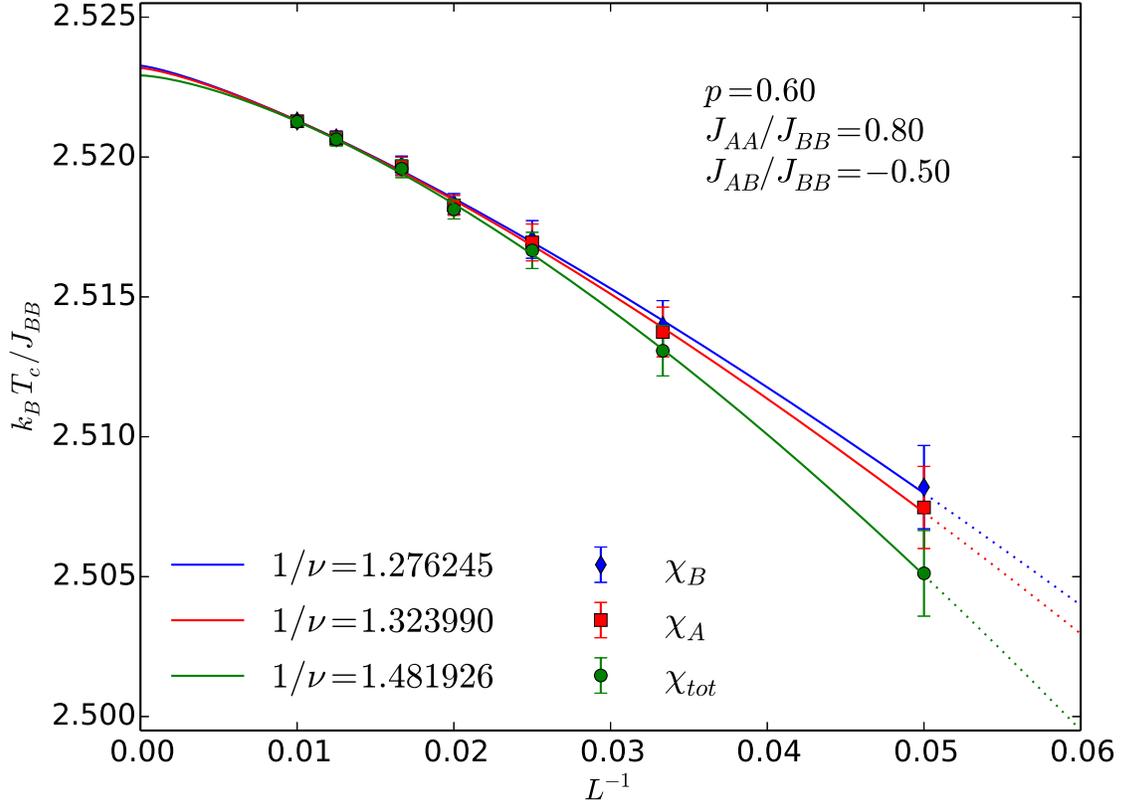}
\caption{
\label{fig:fit:tc}
Dimensionless effective critical temperature $k_BT_c(L)/J_{BB}$ as a function of $L^{-1}$
for $p=0.60$, $J_{AA}/J_{BB}=0.80$, and $J_{AB}/J_{BB}=-0.50$.
The symbols are $T_c(L)$ estimates made by locating the \emph{maxima} of the magnetic susceptibilities
$\chi_\tot$ (circles), $\chi_A$ (squares), and  $\chi_B$ (diamonds) for different system sizes.
The solid lines are fits performed with Eq. \eqref{eq:FSS:tc} for $L_{\smin}\leq L\leq 100$
for the values of $1/\nu$ which minimize the $\chi^2/n_{DOF}$ for each case.
The dotted lines are extrapolations of those fits for $L<L_{\smin}$.
}
\end{center}
\end{figure}

In Fig. \ref{fig:fit:tc} we show examples of these fits of the pseudo-critical temperatures
obtained from the maxima of the magnetic susceptibilities
for $p=0.50$, $J_{AA}/J_{BB}=0.80$, and $J_{AB}/J_{BB}=-0.50$.
In this figure we show the fits using the values of $L_\smin$ and $\nu$ that minimize $\chi^2/n_{DOF}$.
We note that this method gives a very small statistical error for $T_c$, even negligible in some cases,
but it is important to point out that this error is underestimated when compared to the actual error,
obtained through a true non-linear fit.
In order to obtain a more realistic error bar, we follow the criterion used in Ref. \onlinecite{diaz2016monte},
in which the values obtained from fits that give $\chi^2/n_{DOF}$ up to $20\%$ larger than the minimum
are considered in the statistical analysis.

It is also clear from Fig. \ref{fig:fit:tc}
that the values of $\nu$ obtained through this process are imprecise and unreliable,
however, it is worth stressing that it is not our goal to obtain a precise description
of the critical behavior for the model.
Therefore, the values of $1/\nu$ presented in Fig. \ref{fig:fit:tc}
serve only as an ``effective exponent'' used to achieve a good estimate of $T_c$.

\subsection{Determining $T_{comp}$}\label{sec:tcomp}

The compensation point is the temperature $T_{comp}$ where $\m=0$ while
$\mA, \mB\neq 0$, as seen in Figs. \ref{fig:magLmax:a} and \ref{fig:mags:a}.
In order to estimate that temperature we perform simulations for a range of temperatures
around $T_{comp}$ and obtain the $\m$ values as a continuous function of $T$ using the multiple-histogram method,
as show in Fig. \ref{fig:mhist:mag} for the total magnetization
of a system with $p=0.80$, $J_{AA}/J_{BB}=0.65$, $J_{AB}/J_{BB}=-0.80$, and several system sizes $L$ from 20 to 100.
The root of $\langle m_\tot(T)\rangle$ is then located using Brent's method \cite{artigo:brent1973}.
The process is repeated for $N_s$ configurations and the final value and associated error
of $T_{comp}$ are determined as discussed in Sec. \ref{simulations}.

To obtain a final estimate of $T_{comp}$, it is necessary to combine the estimates for different system sizes.
In Fig. \ref{fig:mags:a} we see that the different $T_{comp}(L)$ are fairly close to one another.
However, it is clear from Fig. \ref{fig:mhist:mag} that the smaller lattices provide somewhat inconsistent results.
Fig. \ref{fig:fit:tcomp} shows the size dependence of the compensation temperature estimates obtained from the
same data in Fig. \ref{fig:mhist:mag}. We can see that, as $L$ increases,
the compensation temperature approaches a fixed value.
As the compensation effect is not a critical phenomenon, we have no \emph{a priori}
reason to expect a particular expression for the dependence of $T_{comp}$ on $L$.
Based on the form of the curve, however, we propose a power law behavior similar to the one obeyed by the critical temperature:
\begin{equation}
	\label{eq:tcomp1}
	T_{comp}(L) = a+bL^{-\varepsilon},
\end{equation}
where $a$, $b$, and $\varepsilon$ are parameters to be determined in the fitting process.
Also similar to the critical temperature, we lack the statistical resolution to determine the three parameters independently;
thus, to obtain $T_{comp}$, we resort to the same fitting method described in Sec. \ref{sec:tc} to determine $T_c$.

An alternative procedure is employed: we also fit our data to
\begin{equation}
	\label{eq:tcomp0}
	T_{comp}(L) = a = \mbox{constant},
\end{equation}
for $L>L_{\smin}$, which corresponds to averaging the different compensation temperatures
considering only the values of $L$ after the $T_{comp}(L)$ curve has approximately converged.
The value of $L_{\smin}$ is also determined by minimizing the $\chi^2/n_{DOF}$ of the fit.
The latter method corresponds to the same one used in Ref. \onlinecite{diaz2016monte}.

\begin{table}[h]
\caption{Results of the fits of the size dependence of $T_{comp}$ for $p=0.80$, $J_{AA}/J_{BB}=0.65$, and $J_{AB}/J_{BB}=-0.80$,
using Eqs. \eqref{eq:tcomp1} and \eqref{eq:tcomp0} for different values of $L_{\smin}$.
The numbers in parentheses are the statistical errors given by the least-square fit.
These fits were made using the data presented in Fig. \ref{fig:fit:tcomp}.
}
\label{tab:tcomp}
\begin{tabular}{l|llll|ll}
	\hline\hline
	 & \multicolumn{4}{|l|}{Fits to $T_{comp}(L)=a+bL^{-\varepsilon}$ (Eq. \eqref{eq:tcomp1})} & \multicolumn{2}{l}{Fits to $T_{comp}(L)=a$ (Eq. \eqref{eq:tcomp0})} \\
	\hline
	$L_{\smin}$ & $a$ & $b$ & $\varepsilon$ & $\chi^2/n_{DOF}$ & $a$ & $\chi^2/n_{DOF}$ \\
	\hline
	20  & $3.0669825087(7)$ & $2(3)\times 10^3$                 & $4.35258009946$  & $1.8\times 10^{-2}$ & $3.06702996(1)$   & $3.4\times 10^{-1}$\\
	30  & $3.0669845581(6)$ & $4(6)\times 10^4$                 & $5.11999990708$  & $1.5\times 10^{-2}$ & $3.067011268(3)$  & $6.6\times 10^{-2}$\\
	40  & $3.0669505972(5)$ & $1.60381821004(5)\times 10^{-15}$ & $-5.32766666961$ & $3.1\times 10^{-3}$ & $3.0669965247(2)$ & $5.4\times 10^{-3}$\\
	50  & $3.0669457156(7)$ & $2.057928392(1)\times 10^{-14}$   & $-4.78753257781$ & $3.8\times 10^{-3}$ & $3.0669970562(3)$ & $6.4\times 10^{-3}$\\
	60  & $3.066935560(1)$  & $1.10365686(4)\times 10^{-12}$    & $-3.94933526946$ & $4.9\times 10^{-3}$ & $3.0669983378(3)$ & $7.7\times 10^{-3}$\\
	70  & $3.066944621(3)$  & $5.11588460(2)\times 10^{-14}$    & $-4.59247329163$ & $7.4\times 10^{-3}$ & $3.0670022149(4)$ & $8.6\times 10^{-3}$\\
	\hline\hline
\end{tabular}
\end{table}

Table \ref{tab:tcomp} shows the results of the fits presented in Fig. \ref{fig:fit:tcomp}
with Eqs. \eqref{eq:tcomp1} and \eqref{eq:tcomp0} for different values of $L_{\smin}$.
We see that values of $a$ obtained from each method
are all consistent, irrespective of $\varepsilon$ or $L_{\smin}$.
At first glance, we could be tempted to choose Eq. \eqref{eq:tcomp1} with $L_{\smin}=40$
as the best fit, based on the value of $\chi^2/n_{DOF}$, and use it to obtain the final estimate of $T_{comp}$.
Nonetheless, we note that the fit with Eq. \eqref{eq:tcomp1} cannot be used to determine $T_{comp}$
for any case from Tab. \ref{tab:tcomp} where $L_{\smin}\geq 40$,
given that $\varepsilon<0$ contradicts our assumption that $T_{comp}(L)$ approaches $T_{comp}$ asymptotically
and, in these cases, the limit used to determine the compensation point,
i. e., $T_{comp}=\lim_{L\rightarrow\infty}T_{comp}(L)$, does not exist.
The fact that these particular values of $a$ are so close to the other $T_{comp}$
estimates is a consequence of the small values obtained for parameter $b$.
In fact, in these cases, the term $bL^{-\varepsilon}$ is less than $10^{-4}$, at least for $L\leq 100$.
This means that the dependence with $L$ of Eq. \eqref{eq:tcomp1} becomes irrelevant
to the fitting process, i. e., these results become equivalent to a fit using Eq. \eqref{eq:tcomp0}.
On the other hand, if we impose $\varepsilon>0$ we obtain the best fits for $\varepsilon$ very close to zero ($\varepsilon\approx 10^{-6}$)
for $L_{\smin}\geq 40$, which is consistent with a fit using Eq. \eqref{eq:tcomp0}.
Yet, the fit with such a small value of $\varepsilon$ produces imprecise estimates for $a$ and $b$,
with errors that reach up to three orders of magnitude higher than the parameters themselves.

These results are consistent with the fact that the compensation phenomenon is not in any way related to criticality.
Hence, the observables near the compensation point are not supposed to exhibit a power law scaling behavior.
Consequently, the method using Eq. \eqref{eq:tcomp0} turns out to be more adequate to determine $T_{comp}$,
besides being simpler and more robust.
To estimate the final error bars we follow the same procedure used in Ref. \onlinecite{diaz2016monte},
i. e., we combine the error obtained in the fitting process with both the errors obtained for $N_s$ samples
and for a single sample via the blocking method.

\section{Results and Discussion}\label{results}

\begin{figure}[h]
\begin{center}
\subfigure[With compensation.\label{fig:magLmax:a}]{
\includegraphics[width=\subfigwidth]{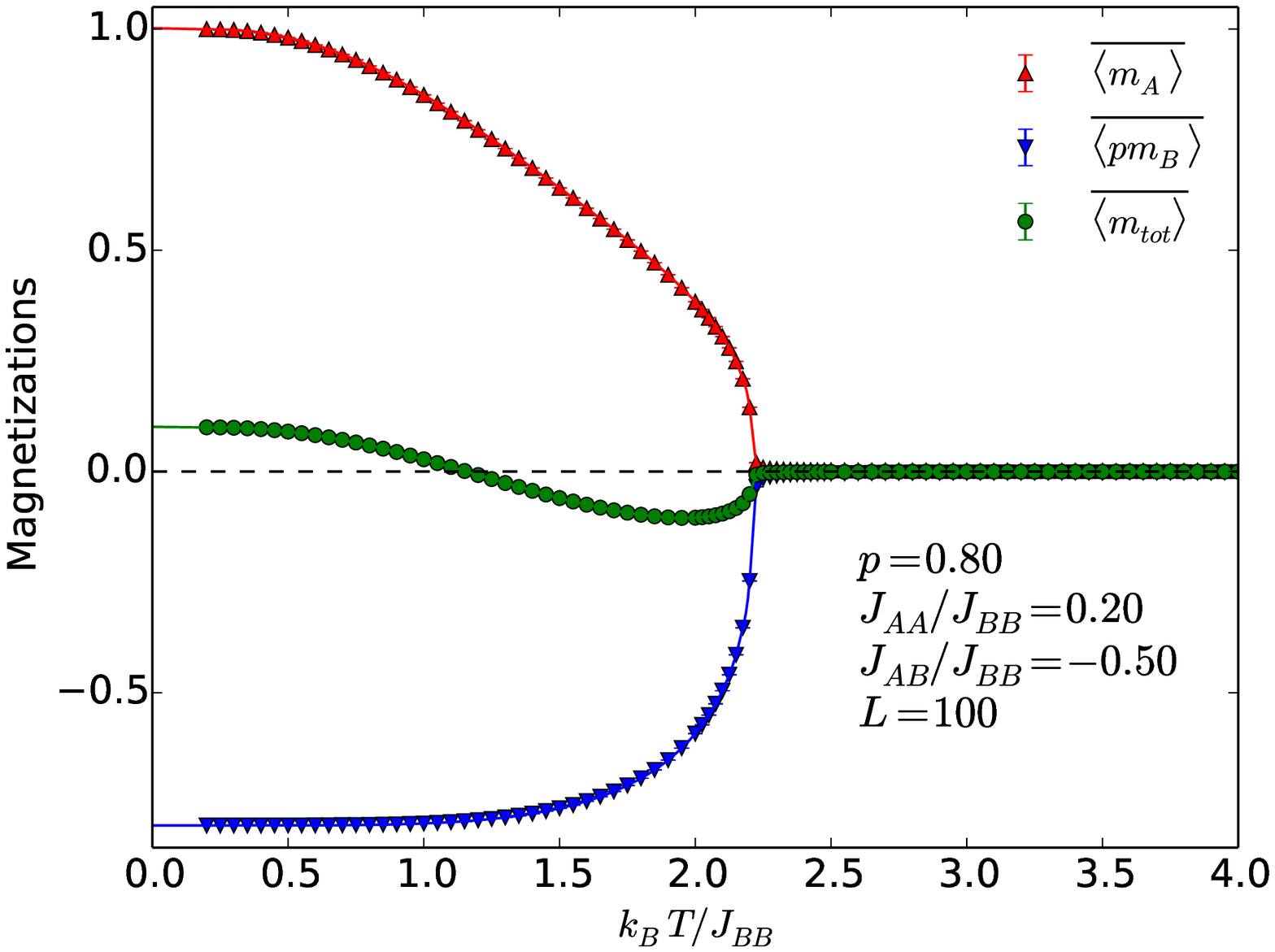}
}
\subfigure[Without compensation.\label{fig:magLmax:b}]{
\includegraphics[width=\subfigwidth]{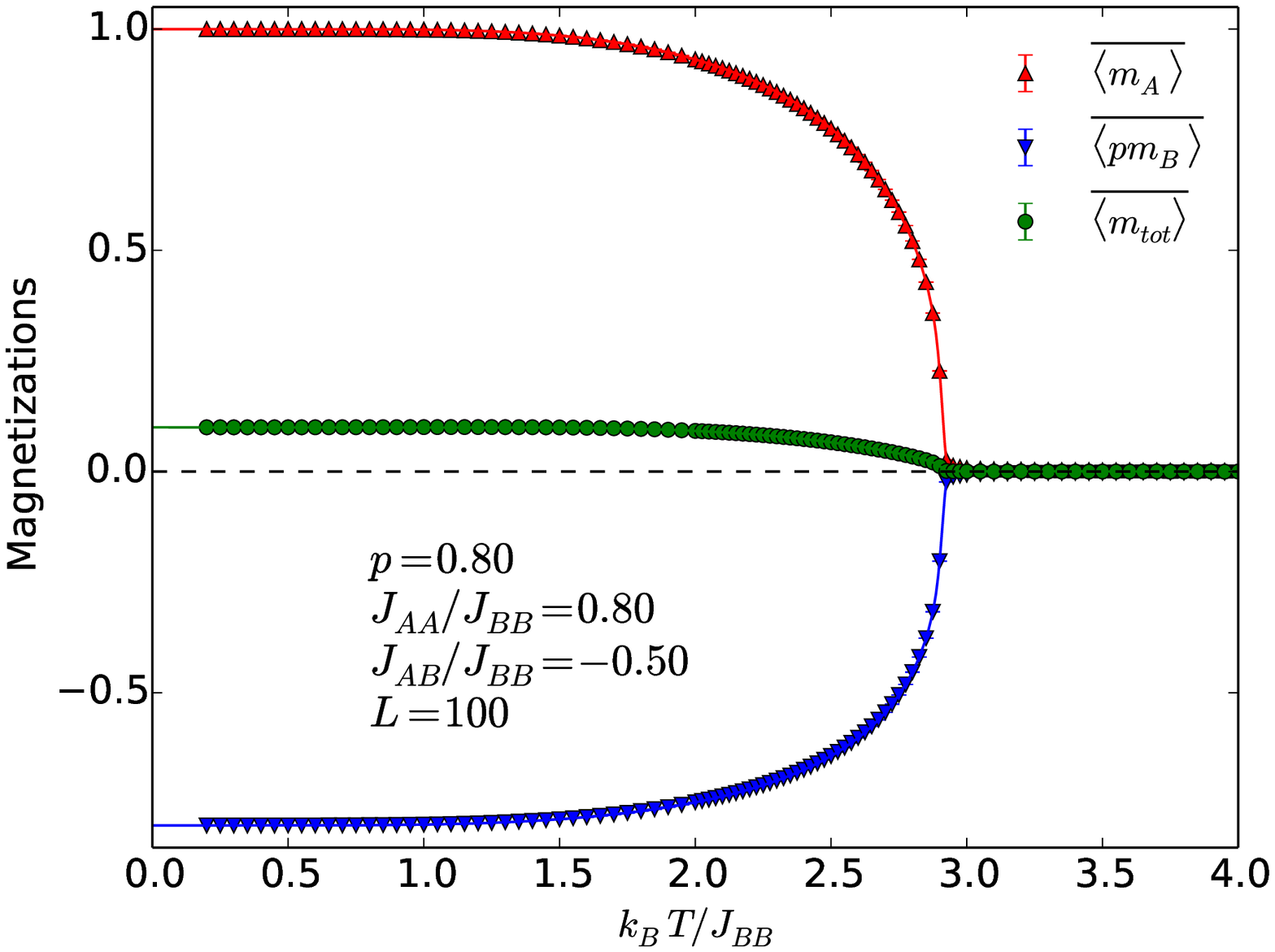}
}
\caption{
\label{fig:magLmax}
Plane magnetizations, $\overline{\mA}$ and $\overline{\pmB}$, and total magnetization $\overline{\m}$
as a function of the dimensionless temperature $k_BT/J_{BB}$ for $p=0.80$, $J_{AB}/J_{BB}=-0.50$ and $L=100$.
Figure (a), for $J_{AA}/J_{BB}=0.20$, shows a compensation temperature $T_{comp}$
such that $\overline{\m}=0$ and $0<T_{comp}<T_c$
whereas figure (b), for $J_{AA}/J_{BB}=0.80$, shows no compensation effect.
The symbols correspond to the data and the solid lines are cubic spline interpolations just to guide the eye.
The error bars are smaller than the symbols.
}
\end{center}
\end{figure}

\begin{figure}[h]
\begin{center}
\subfigure[With compensation.\label{fig:mags:a}]{
\includegraphics[width=\subfigwidth]{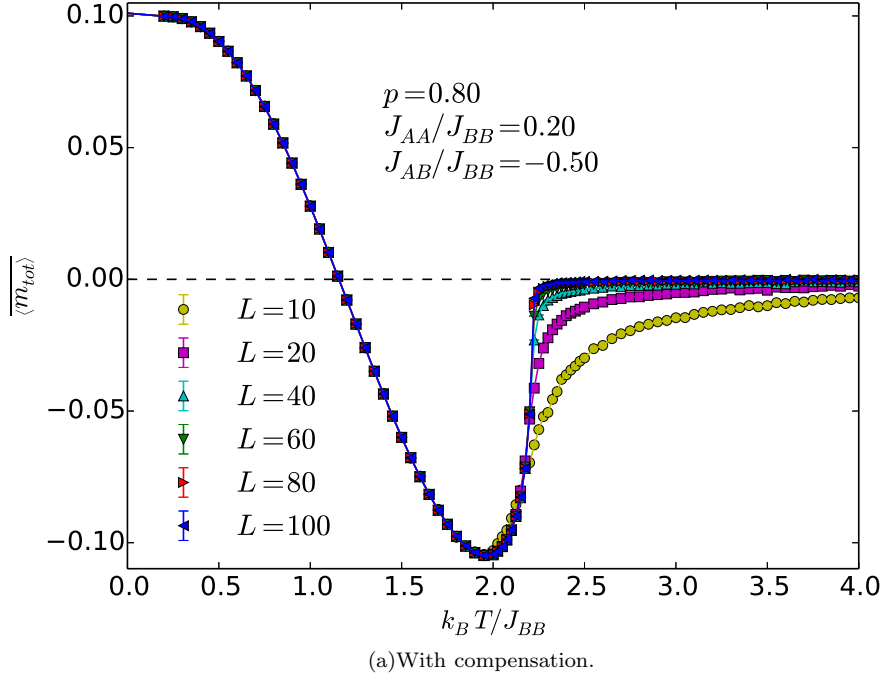}
}
\subfigure[Without compensation.\label{fig:mags:b}]{
\includegraphics[width=\subfigwidth]{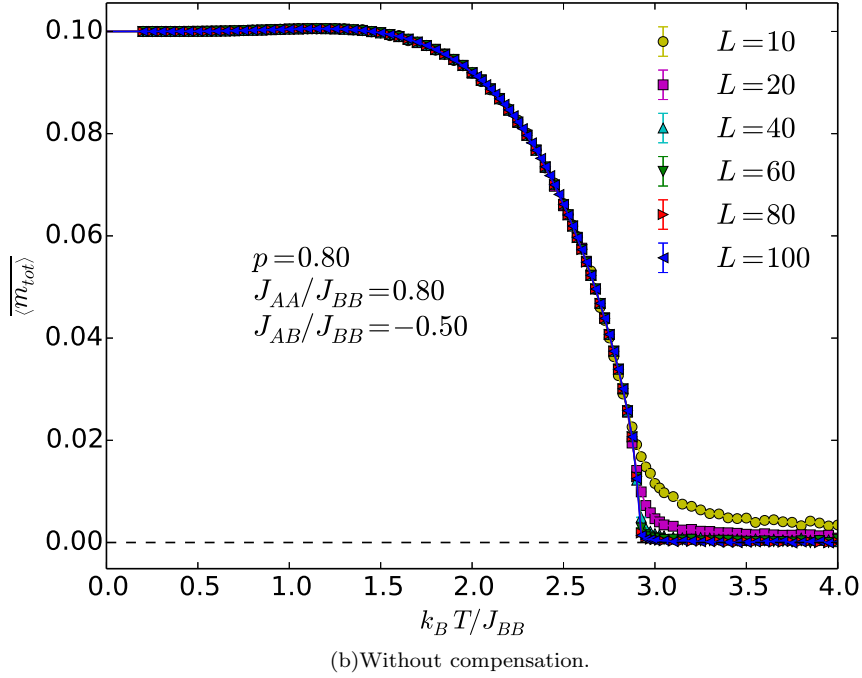}
}
\caption{
\label{fig:mags}
Total magnetization $\overline{\m}$ as a function of the the dimensionless temperature $k_BT/J_{BB}$
for $p=0.80$, $J_{AB}/J_{BB}=-0.50$, and several values of system size $L$.
Figure (a), for $J_{AA}/J_{BB}=0.20$, shows a compensation temperature $T_{comp}$
such that $\overline{\m}=0$ and $0<T_{comp}<T_c$
whereas figure (b), for $J_{AA}/J_{BB}=0.80$, shows no compensation effect.
The symbols correspond to the data and the solid lines are cubic spline interpolations just to guide the eye.
The error bars are smaller than the symbols.
}
\end{center}
\end{figure}

\begin{figure}[h]
\begin{center}
\includegraphics[width=\figwidth]{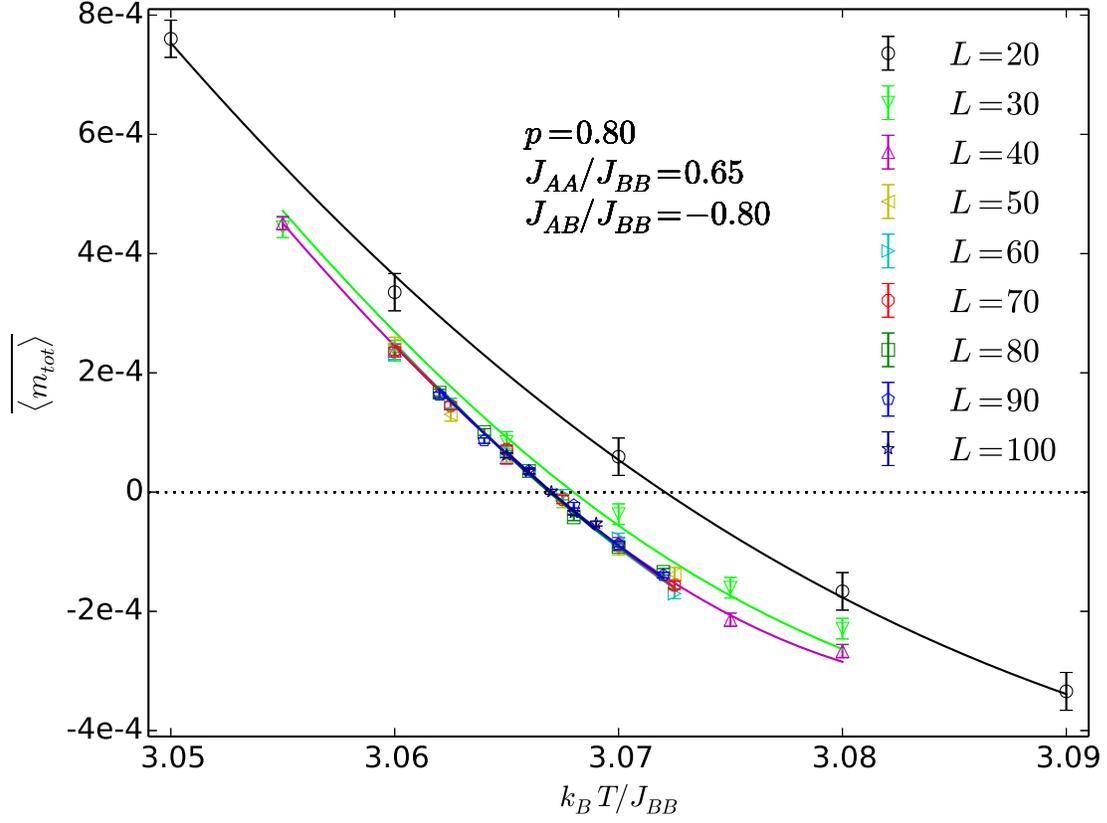}
\caption{
\label{fig:mhist:mag}
Total magnetization as a function of the dimensionless temperature $k_BT/J_{BB}$
for  $p=0.80$, $J_{AA}/J_{BB}=0.65$, $J_{AB}/J_{BB}=-0.80$, and linear lattice sizes $L$ ranging from 20 to 100.
The symbols correspond to simulation data and the solid lines were obtained using the multiple histogram method.
}
\end{center}
\end{figure}

\begin{figure}[h]
\begin{center}
\includegraphics[width=\figwidth]{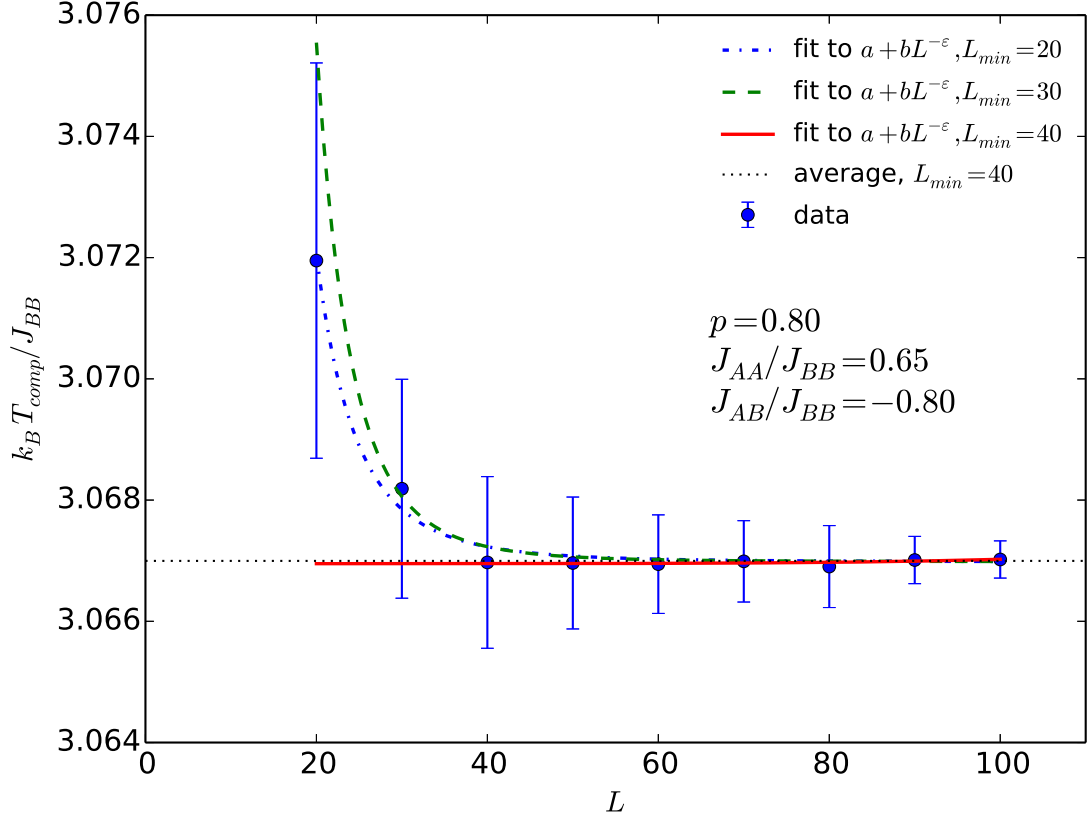}
\caption{
\label{fig:fit:tcomp}
Dimensionless compensation temperature $k_BT_{comp}(L)/J_{BB}$ as a function of $L$
for $p=0.80$, $J_{AA}/J_{BB}=0.65$, and $J_{AB}/J_{BB}=-0.80$.
The symbols are estimates made by locating the zero of the total magnetization for different system sizes.
The lines are fits performed with Eq. \eqref{eq:tcomp1} for the values of $\varepsilon$ which minimize the $\chi^2/n_{DOF}$
for each case or with Eq. \eqref{eq:tcomp0} for $L_{\smin}\leq L\leq 100$ with $L_{\smin}=40$,
which is the value that minimizes the $\chi^2/n_{DOF}$ of this type of fit.
}
\end{center}
\end{figure}

The system studied in this work has three parameters in its Hamiltonian,
namely the concentration of magnetic sites $p$, and the ratios $J_{AA}/J_{BB}$ and $J_{AB}/J_{BB}$,
which represent, respectively, the asymmetry between type-\textbf{A} and type-\textbf{B} intraplanar couplings
and the relative strength of interplanar coupling in relation to the type-\textbf{B} intraplanar coupling.
All these parameters play important roles in determining the behavior of the system.
Our goal is to outline the contribution of each parameter to the presence or absence
of the compensation phenomenon.
To that end we map out the regions of the parameter space for which the system has
a compensation point, as seen in Figs. \ref{fig:magLmax:a} and \ref{fig:mags:a},
and the regions for which the compensation effect does not take place,
as seen in Figs. \ref{fig:magLmax:b} and \ref{fig:mags:b}.

\begin{figure}[h]
\begin{center}
\includegraphics[width=\figwidth]{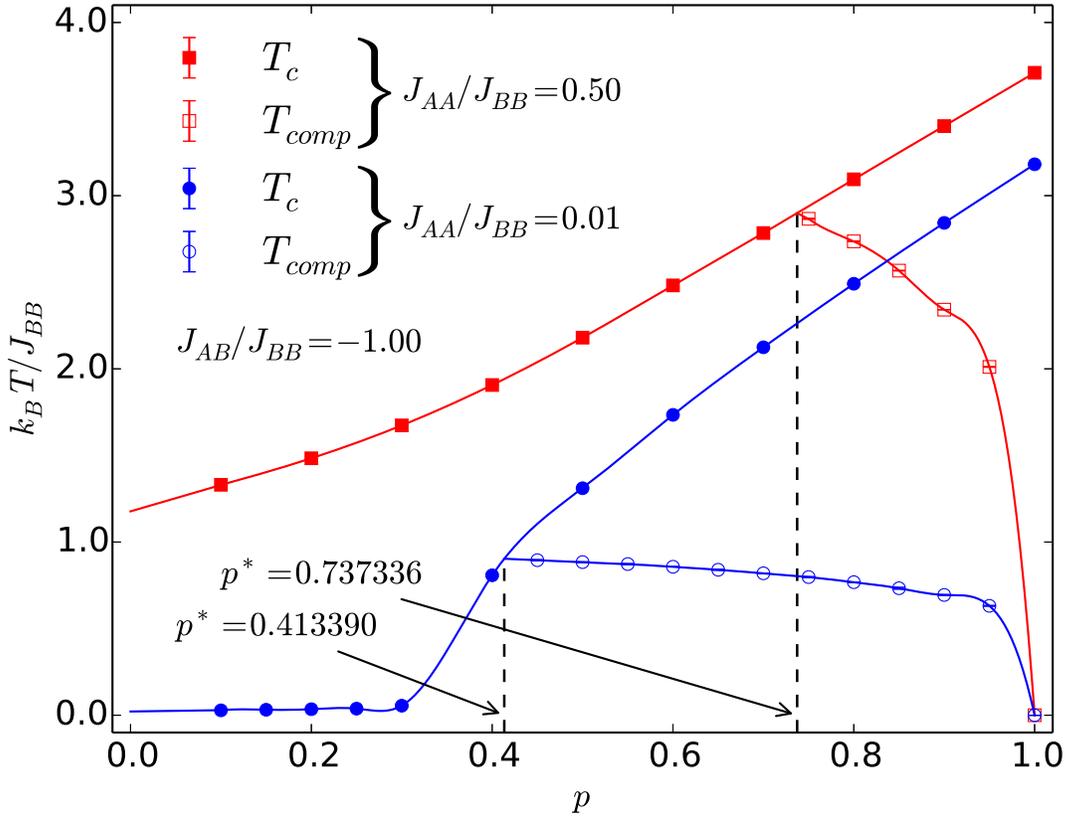}
\caption{
\label{fig:Tvsp}
Critical temperatures $T_c$ (filled symbols)
and compensation temperatures $T_{comp}$ (empty symbols) as functions of the spin concentration $p$
for $J_{AB}/J_{BB}=-1.00$ in both cases: $J_{AA}/J_{BB}=0.01$ (circles) and $J_{AA}/J_{BB}=0.50$ (squares).
The solid lines are either cubic spline interpolations or linear extrapolations just to guide the eye.
The vertical dashed lines mark the characteristic concentration $p^\ast$ where $T_c=T_{comp}$ and below which there is no compensation.
Where the error bars are not visible, they are smaller than the symbols.
}
\end{center}
\end{figure}

We start our analysis by fixing the values of $J_{AA}/J_{BB}$ and $J_{AB}/J_{BB}$
and looking at the dependence of both the critical and compensation temperatures with $p$,
as seen in Fig. \ref{fig:Tvsp}, where we plot $T_c$ and $T_{comp}$ as functions of $p$
for $J_{AB}/J_{BB}=-1.00$ and for both $J_{AA}/J_{BB}=0.01$ and $J_{AA}/J_{BB}=0.50$.
The solid lines are either cubic spline interpolations or linear extrapolations just to guide the eye.
The vertical dashed lines mark the characteristic concentration $p^\ast$ where 
the critical temperature and compensation temperature curves meet and below which there is no compensation.

\begin{figure}[h]
\begin{center}
\includegraphics[width=\figwidth]{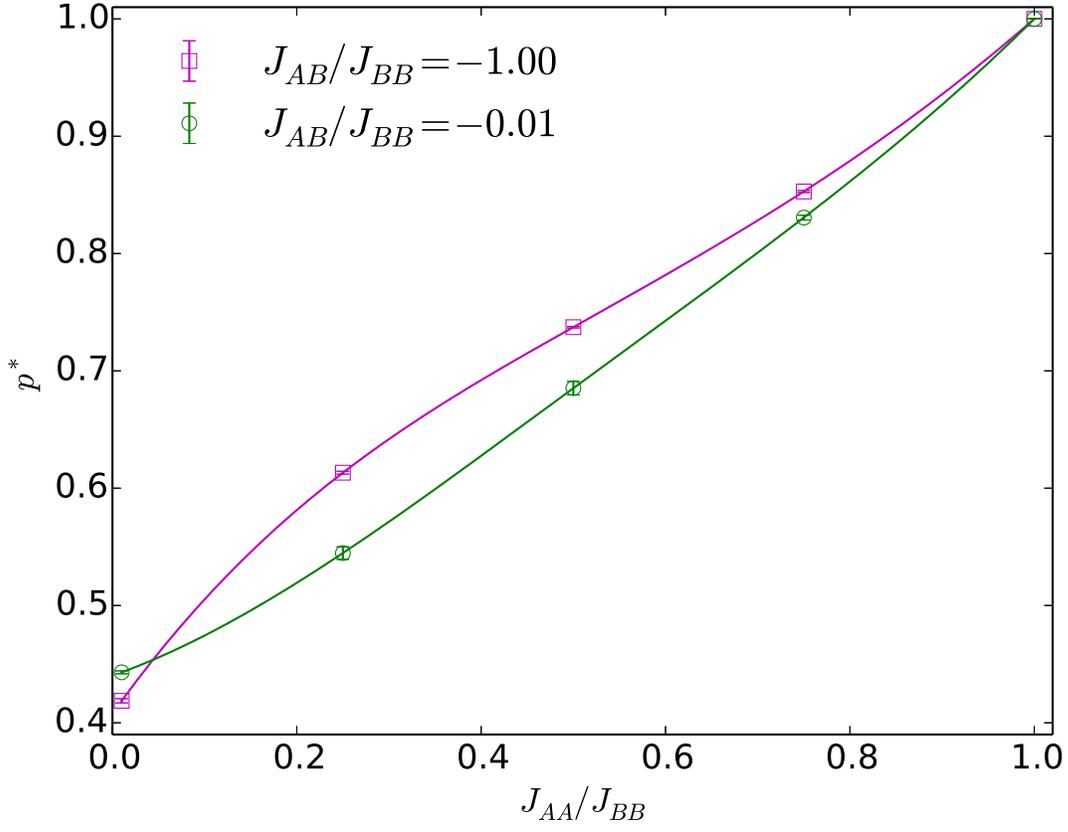}
\caption{
\label{fig:pvsJAA}
Characteristic concentration $p^\ast$, below which there is no compensation,
as a function of the ratio $J_{AA}/J_{BB}$ for $J_{AB}/J_{BB}=-0.01$ (circles) and $J_{AB}/J_{BB}=-1.00$ (squares).
The solid lines are cubic spline interpolations just to guide the eye.
Where the error bars are not visible, they are smaller than the symbols.
}
\end{center}
\end{figure}

The behavior depicted in Fig. \ref{fig:Tvsp} for the multilayer (3d) system is qualitatively
the same as displayed by the bilayer (2d) system
for both pair approximation \cite{balcerzak2014ferrimagnetism} and Monte Carlo \cite{diaz2016monte}.
This can be seen by comparing our Fig. \ref{fig:Tvsp} to Fig. 2 in reference \onlinecite{balcerzak2014ferrimagnetism}
and Fig. 7 in reference \onlinecite{diaz2016monte}, since all where made using the same values for the Hamiltonian parameters.
As expected, the critical temperatures obtained for the 3d system
are consistently higher than those reported for the 2d one using the same approximation \cite{diaz2016monte}.
In all cases, though, we see that $p^\ast$ is higher for $J_{AA}/J_{BB}=0.50$ than it is for $J_{AA}/J_{BB}=0.01$,
which indicates that $p^\ast$ should increase as the strength of the \textbf{A} intraplanar coupling increases.
This is expected, since, for high $J_{AA}/J_{BB}$, the magnetization of planes \textbf{B} will never
be the same (in absolute value) as the magnetization of planes \textbf{A} for small $p$.
We are able to confirm this trend, for both cases of a weak and a strong interplanar coupling,
in Fig. \ref{fig:pvsJAA},
where we show $p^\ast$ as a function of the ratio $J_{AA}/J_{BB}$ for
$J_{AB}/J_{BB}=-0.01$  and $J_{AB}/J_{BB}=-1.00$. In both cases we see that $p^\ast$
increases monotonically as $J_{AA}/J_{BB}$ increases.

\begin{figure}[h]
\begin{center}
\includegraphics[width=\figwidth]{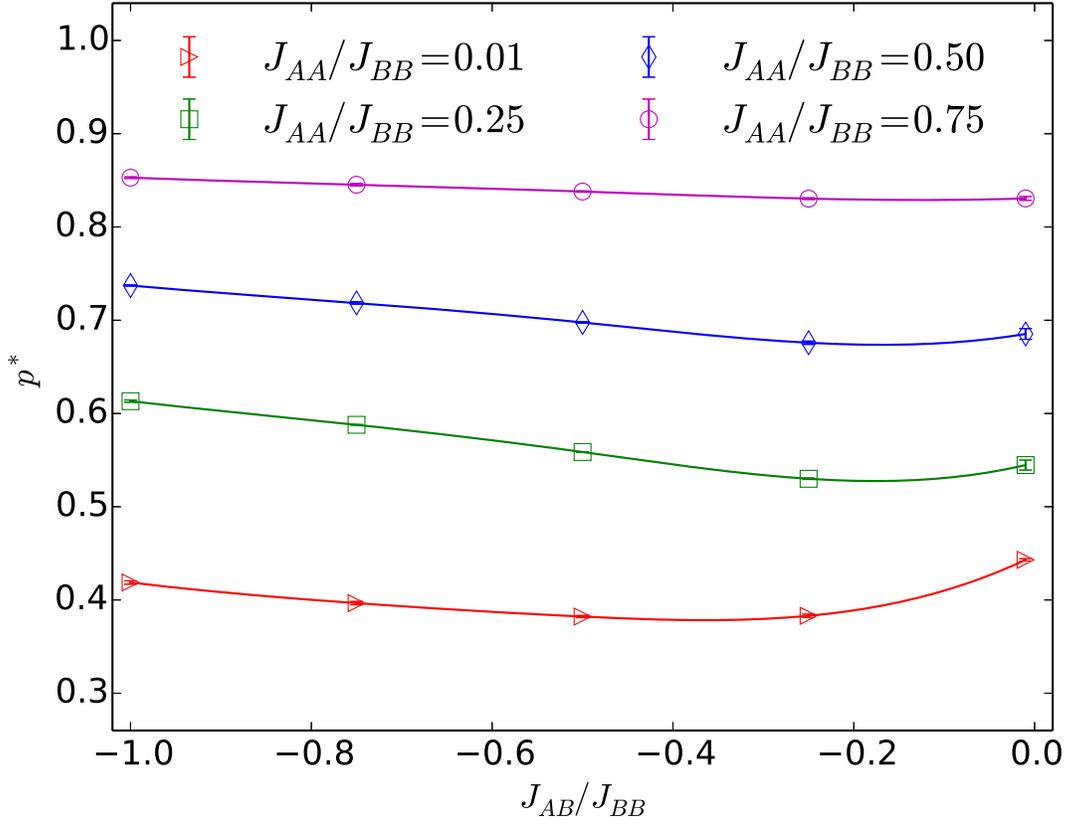}
\caption{
\label{fig:pvsJAB}
Characteristic concentration $p^\ast$, below which there is no compensation,
as a function of the ratio $J_{AB}/J_{BB}$ for several values of $J_{AA}/J_{BB}$.
The solid lines are cubic spline interpolations just to guide the eye.
Where the error bars are not visible, they are smaller than the symbols.
}
\end{center}
\end{figure}

In Fig. \ref{fig:pvsJAB} we have the behavior of $p^\ast$
as a function of the ratio $J_{AB}/J_{BB}$ for several values of $J_{AA}/J_{BB}$.
Curiously, $p^\ast$ is not monotonic as is the case for the 2d system \cite{diaz2016monte}.
Figure 9 in Ref. \onlinecite{diaz2016monte} shows the same $p^\ast \times J_{AB}/J_{BB}$ diagram
for the bilayer system for the particular case of $J_{AA}/J_{BB}=0.50$
and, comparing the latter figure to the 3d results
we see that the values of $p^\ast$ for the 3d system are consistently lower
than those for the 2d one.
However, this difference gets smaller as $|J_{AB}/J_{BB}|$ gets weaker
($p^\ast$ for the multilayer is $\approx 8\%$ lower than value for the bilayer for $J_{AB}/J_{BB}=-1.00$
and it is $\approx 3\%$ lower for $J_{AB}/J_{BB}=-0.01$),
which is consistent with the fact that as $J_{AB}/J_{BB}\rightarrow 0$
the behaviors of both multilayer and bilayer systems cross over to
the 2d behavior of non-interacting planes \textbf{A} and \textbf{B}.
We also notice that the values of $p^\ast$ are less dependent on the values $J_{AB}/J_{BB}$ for 3d than for 2d.
For instance, for $J_{AA}/J_{BB}=0.50$, the percentile increase in $p^\ast$ as $|J_{AB}/J_{BB}|$ increases from $0.01$ to $1.00$
is $\approx 13\%$ for the bilayer, whereas for the multilayer this increase is only $\approx 7.6\%$.
Even if we consider the maximum percentile increase (which for 2d remains $\approx 13\%$ for its monotonic behavior),
we have, for the multilayer, at most an increase of $\approx 10\%$ from the minimum at about $J_{AB}/J_{BB}=-0.20$ to the
maximum at $J_{AB}/J_{BB}=-1.00$.

%

\begin{figure}[h]
\begin{center}
\subfigure[$p=0.70$\label{fig:TvsJAA:a}]{
\includegraphics[width=\subfigwidth]{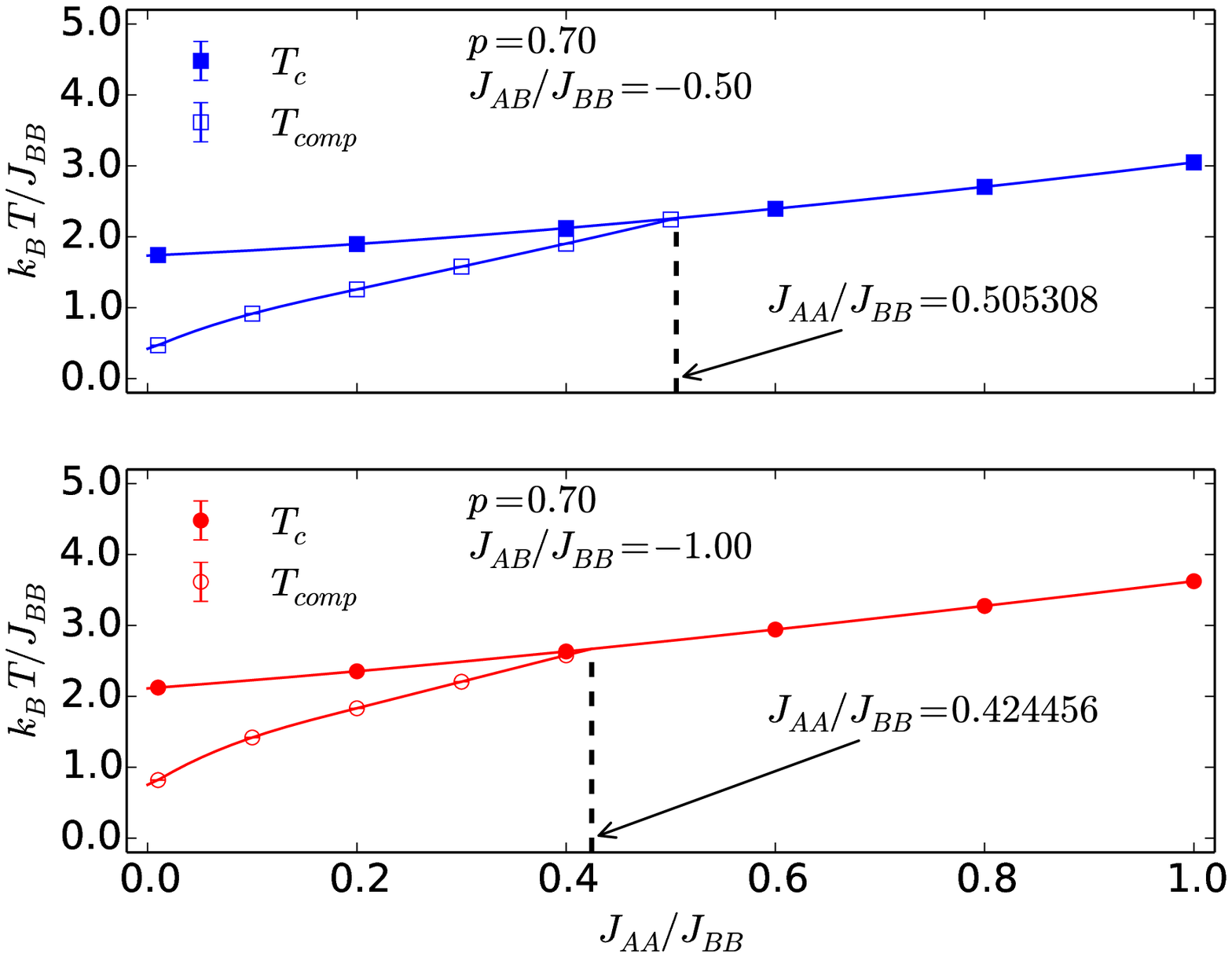}
}
\subfigure[$p=0.90$\label{fig:TvsJAA:b}]{
\includegraphics[width=\subfigwidth]{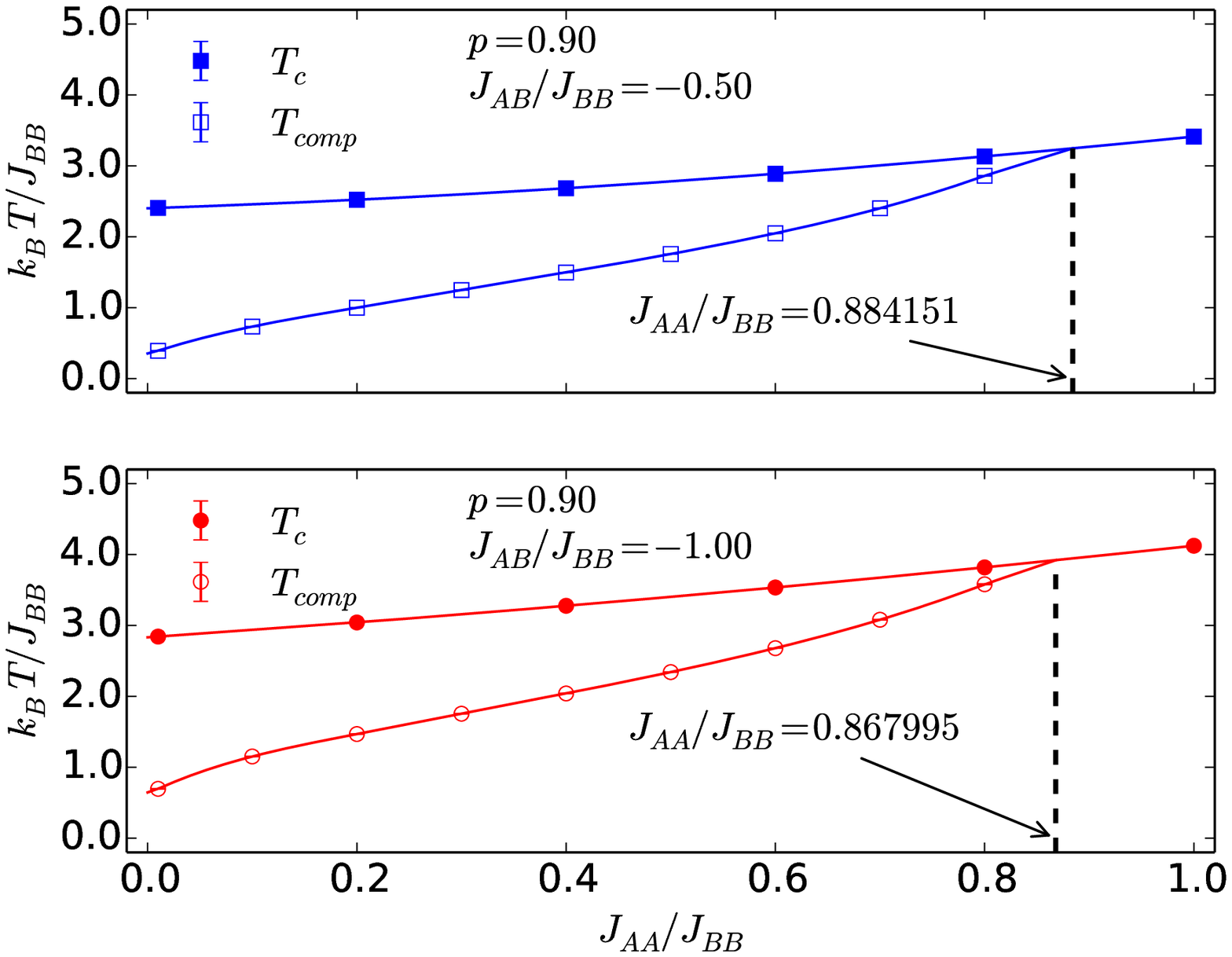}
}
\caption{
\label{fig:TvsJAA}
Critical temperatures $T_c$ (filled symbols)
and compensation temperatures $T_{comp}$ (empty symbols) as functions of the ratio $J_{AA}/J_{BB}$
for $J_{AB}/J_{BB}=-0.50$ and $J_{AB}/J_{BB}=-1.00$ in both cases: (a) $p=0.70$ and (b) $p=0.90$.
The solid lines are either cubic spline interpolations or linear extrapolations just to guide the eye.
The vertical dashed lines mark the value of $J_{AA}/J_{BB}$ where $T_c=T_{comp}$ and above which there is no compensation.
Where the error bars are not visible, they are smaller than the symbols.
}
\end{center}
\end{figure}

Next, for fixed values of $p$ and $J_{AB}/J_{BB}$, we can analyze the influence of
the ratio $J_{AA}/J_{BB}$ in the behavior of our multilayer.
In Fig. \ref{fig:TvsJAA}
we plot $T_c$ and $T_{comp}$ as functions of $J_{AA}/J_{BB}$
for $p=0.70$ (Fig. \ref{fig:TvsJAA:a}) and $p=0.90$ (Fig. \ref{fig:TvsJAA:b}),
and fixed $J_{AB}/J_{BB}=-0.50$ and $J_{AB}/J_{BB}=-1.00$ for both cases.
The vertical dashed lines mark the value of $J_{AA}/J_{BB}$ above which there is no compensation for each case.

In order to  contrast the behaviors of 3d and 2d systems,
we can compare Fig. \ref{fig:TvsJAA:a} with Fig. 10 in Ref. \onlinecite{diaz2016monte},
as well as Fig. \ref{fig:TvsJAA:b} with Fig. 11 in the same reference,
since in both pairs of figures we have the same quantities calculated for the same values of fixed parameters.
For instance, for $(p, J_{AB}/J_{BB})=(0.70, -1.00)$,
the value of $J_{AA}/J_{BB}$ where $T_c=T_{comp}$ is $\approx 77\%$ higher for the 3d system,
whereas, for $(p, J_{AB}/J_{BB})=(0.90, -0.50)$, this difference drops to $\approx 10\%$.
Still, in both cases the region of the diagram for which the system has a compensation point is larger for the multilayer.





Also, in both cases we have the critical temperatures for the 3d system
consistently higher than those obtained with MC simulations for the bilayer, as expected.
For $(p, J_{AB}/J_{BB})=(0.70, -1.00)$, the critical temperature for the 3d system is
almost twice as high as the 2d value at $J_{AA}/J_{BB}=0.0$,
while at $J_{AA}/J_{BB}=1.0$, the 3d value is only $37\%$ higher.
For $(p, J_{AB}/J_{BB})=(0.90, -0.50)$,
the 3d system $T_c$ is $26\%$ higher than the 2d one at $J_{AA}/J_{BB}=0.0$
and $25\%$ higher at $J_{AA}/J_{BB}=1.0$.
Still consistent with the 3d to 2d crossover, the difference between the 3d and 2d
critical temperatures is bigger for a stronger interplanar coupling.
However, this difference does not get smaller as $J_{AA}/J_{BB}$
gets weaker, which agrees with the fact that there is no crossover from 3d to 2d as $J_{AA}/J_{BB}\rightarrow 0$.
As a matter of fact, we are able to set $J_{AA}/J_{BB}=0$ and the system is still a multilayer
as long as $p\neq 0$ and $J_{AB}/J_{BB}\neq 0$.

We can also compare our multilayer MC results with the
bilayer PA behavior reported in Ref. \onlinecite{balcerzak2014ferrimagnetism}.
This is made by contrasting
Fig. \ref{fig:TvsJAA:a} with Fig. 6 in the latter reference,
as well as Fig. \ref{fig:TvsJAA:b} with Fig. 7 in the same reference.
After correcting for the difference in temperature scale
(due to the PA using $s_i\pm 1/2$ and MC using $s_i\pm 1$),
as discussed in Ref. \onlinecite{diaz2016monte},
for $(p, J_{AB}/J_{BB})=(0.70, -1.00)$
we have the critical temperature for the 3d system in MC $\approx 90\%$
higher than for the 2d system in the PA at $J_{AA}/J_{BB}=0.0$
and $12\%$ higher at $J_{AA}/J_{BB}=1.0$.
For $(p, J_{AB}/J_{BB})=(0.90, -0.50)$
the critical temperature for the 3d system in MC is $\approx 1.3\%$ lower
than for the 2d system in the PA at $J_{AA}/J_{BB}=0.0$
and $4.7\%$ higher at $J_{AA}/J_{BB}=1.0$.
Again we see that the agreement is better for weaker interplanar couplings as
in this case both systems are closer to a 2d behavior.
It is also worth mentioning that the PA results for the bilayer
are much closer to the MC results for the multilayer than to the MC results for the bilayer.
This is consistent with the fact that mean-field-like approximations tend to
work better in higher dimensions.

\begin{figure}[h]
\begin{center}
\subfigure[$p=0.60$\label{fig:TvsJAB:a}]{
\includegraphics[width=\subfigwidth]{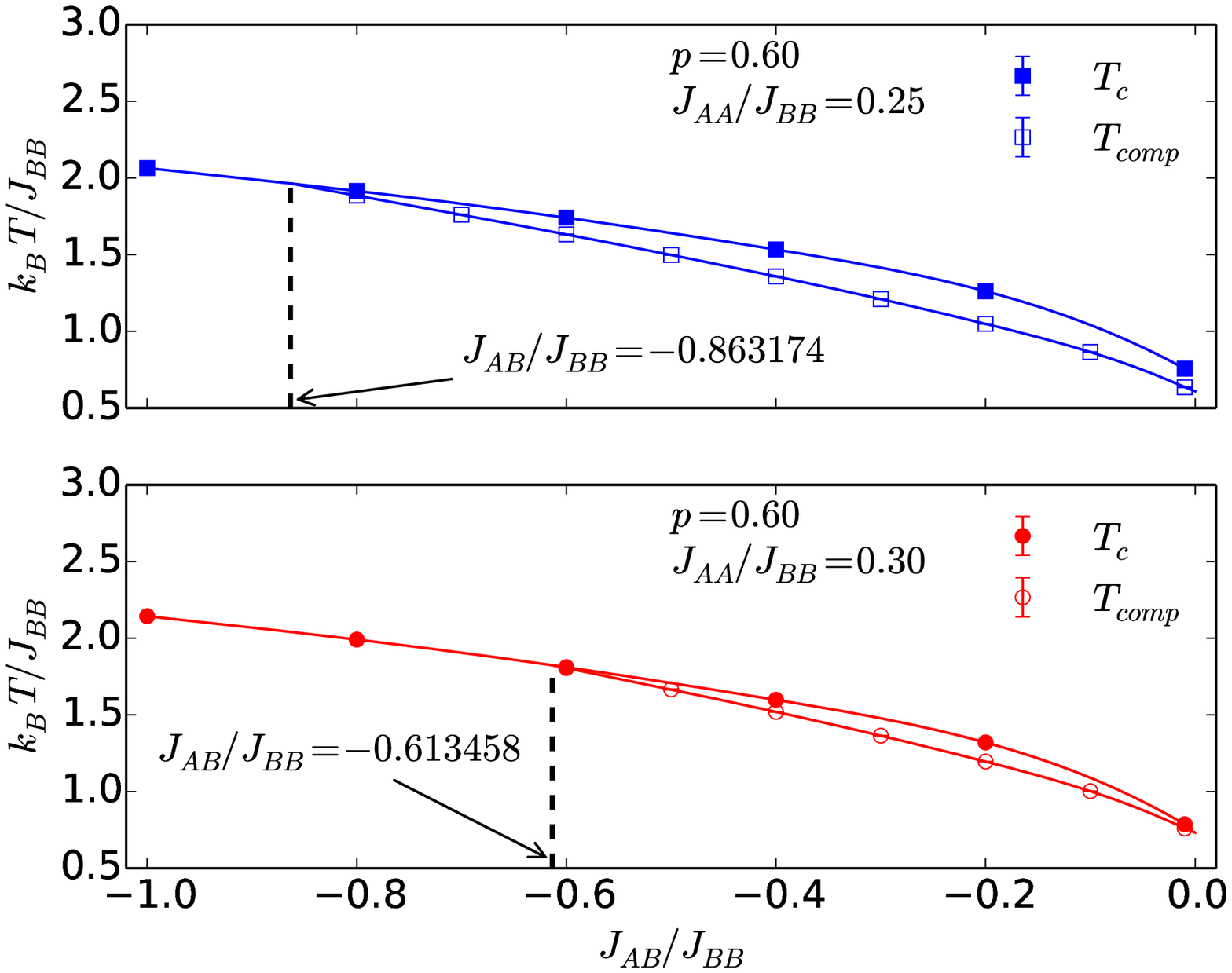}
}
\subfigure[$p=0.70$\label{fig:TvsJAB:b}]{
\includegraphics[width=\subfigwidth]{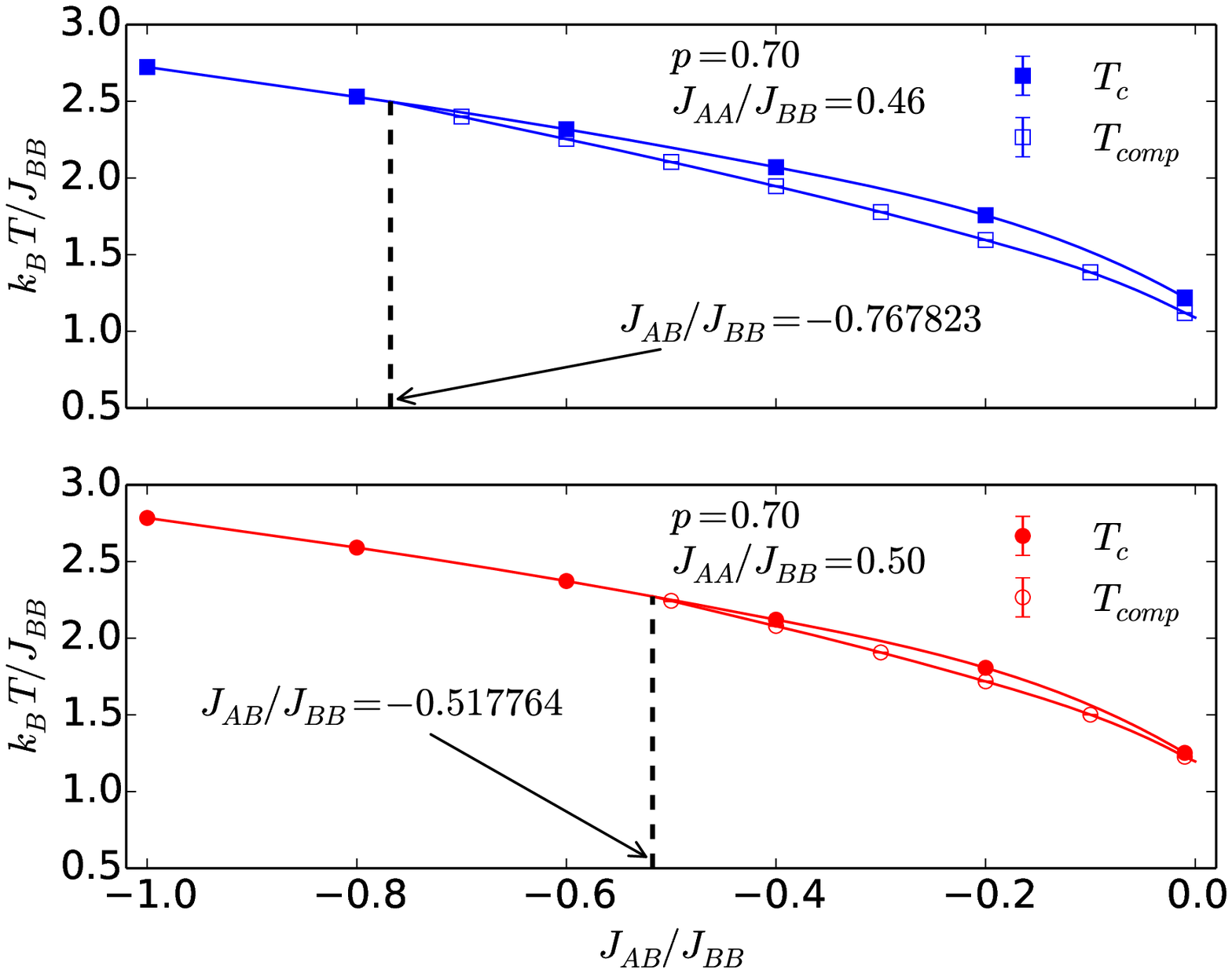}
}
\caption{
\label{fig:TvsJAB}
Critical temperatures $T_c$ (filled symbols)
and compensation temperatures $T_{comp}$ (empty symbols) as functions of the ratio $J_{AB}/J_{BB}$
In (a) we have $p=0.60$ for both $J_{AA}/J_{BB}=0.25$ and $J_{AA}/J_{BB}=0.30$.
In (b) we have $p=0.70$ for both $J_{AA}/J_{BB}=0.46$ and $J_{AA}/J_{BB}=0.50$.
The solid lines are either cubic spline interpolations or linear extrapolations just to guide the eye.
The vertical dashed lines mark the value of $J_{AB}/J_{BB}$ where $T_c=T_{comp}$ and below which there is no compensation.
Where the error bars are not visible, they are smaller than the symbols.
}
\end{center}
\end{figure}

In Fig. \ref{fig:TvsJAB}
we have $T_c$ and $T_{comp}$ as functions of $J_{AB}/J_{BB}$
for $(p, J_{AA}/J_{BB})=(0.60, 0.25), (0.60, 0.30)$ (\ref{fig:TvsJAB:a}) and
$(p, J_{AA}/J_{BB})=(0.70, 0.46), (0.70, 0.50)$ (\ref{fig:TvsJAB:b}).
The vertical dashed lines mark the value of $J_{AB}/J_{BB}$ below which there is no compensation for each case.
Unfortunately no direct comparison can be made between this figure and the bilayer MC results,
since the analogous figure for the bilayer (Fig. 12 in Ref. \onlinecite{diaz2016monte})
does not have the same parameters as ours.
The reason we did not use the same parameter for our figure \ref{fig:TvsJAB}
as those used for Fig. 12 in Ref. \onlinecite{diaz2016monte}
is that both sets of parameters in the latter figure ($(p, J_{AA}/J_{BB})=(0.7, 0.3)$ and $(0.9, 0.8)$)
are in a ferrimagnetic phase where there is always compensation for $-1.0\leq J_{AB}/J_{BB}<0.0$.
Qualitatively, though, it is safe to say that
the $T_c$ and $T_{comp}$ curves do not get further apart as $J_{AB}/J_{BB}\rightarrow 0$
in the multilayer system as it happens in the bilayer,
for both MC \cite{diaz2016monte} and PA \cite{balcerzak2014ferrimagnetism}.
In fact, for fixed $p$ and $J_{AA}/J_{BB}$ we seem to have
$T_c$ and $T_{comp}$ always very close to one another,
and the difference $T_c-T_{comp}$ dos not increase monotonically as $J_{AB}/J_{BB}$ increases,
having a maximum value somewhere between the point where both curves meet and $J_{AB}/J_{BB}=0.0$.

\begin{figure}[h]
\begin{center}
\includegraphics[width=\figwidth]{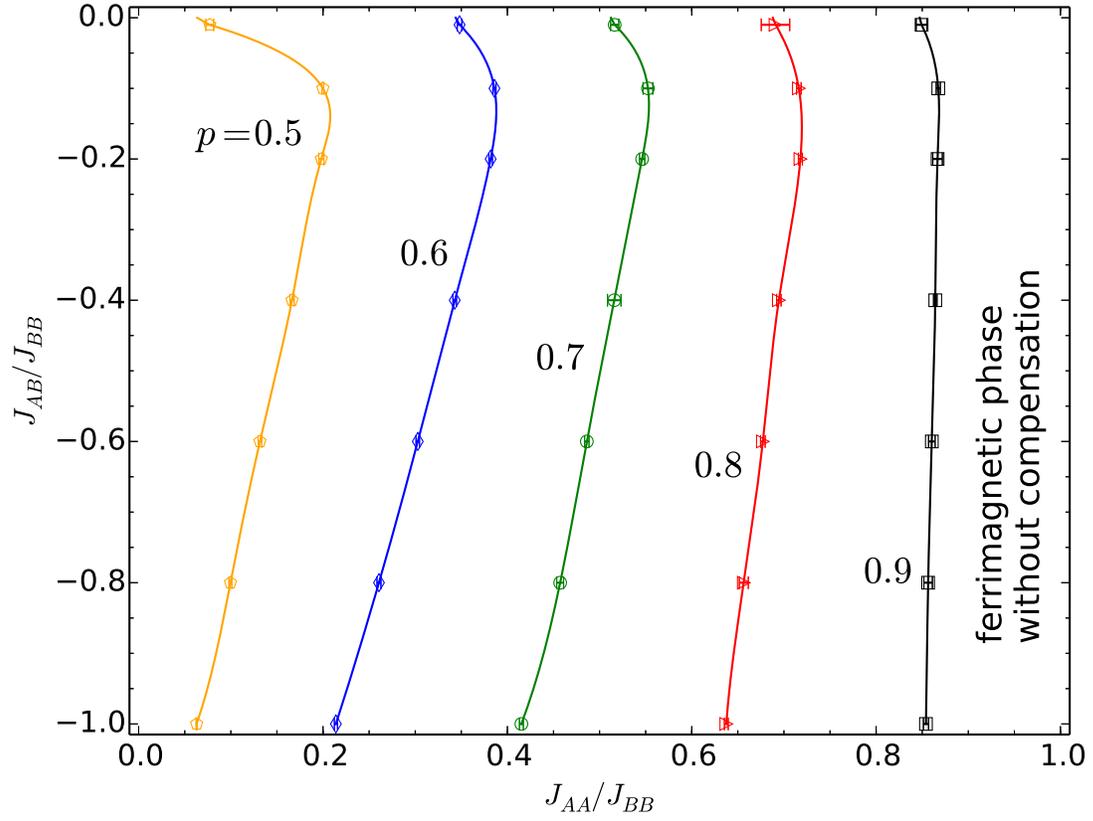}
\caption{
\label{fig:phase}
Phase diagrams for concentrations $p=0.5, 0.6, 0.7, 0.8$, and $0.9$.
The solid lines are either cubic spline interpolations or linear extrapolations just to guide the eye.
For each concentration, the line marks the separation between a ferrimagnetic phase with compensation (to the left)
and a ferrimagnetic phase without compensation (to the right).
}
\end{center}
\end{figure}

Finally, as it follows from the analyzes presented above, we can
divide the parameter space of our Hamiltonian in two distinct regions of interest.
One is a ferrimagnetic phase for which there is no compensation at any temperature
and the second is a ferrimagnetic phase where there is a compensation point
at a certain temperature $T_{comp}$.
We present this results
in Fig. \ref{fig:phase},
where we plot the phase diagrams for concentrations $p=0.5, 0.6, 0.7, 0.8$, and $0.9$.
For each concentration, the line marks the separation between a ferrimagnetic phase with compensation (to the left)
and a ferrimagnetic phase without compensation (to the right).

These diagrams show that the compensation phenomenon is favored
by greater values of $p$, as it is clear that as $p$ approaches $1$, as long as $p\neq 1$,
the area occupied by the ferrimagnetic phase with compensation greatly increases.
It is also clear that the compensation phenomenon is only present
if there is intraplanar coupling asymmetry and, as $p$ decreases,
it is necessary to increase the asymmetry for the phenomenon to occur.
We also notice that, as $p$ increases, the line separating the phases becomes more vertical,
i. e., the presence or absence of compensation becomes less sensitive to the value of $J_{AB}/J_{BB}$.
Nonetheless, on all cases presented it is possible to intersect each curve with an actual vertical line in two points,
although this is easier to see for smaller values $p$.
Therefore, there are sets of values for the parameters for which
we can be in a phase without compensation, decrease $|J_{AB}/J_{BB}|$
to get to the phase with compensation,
but if we keep decreasing $|J_{AB}/J_{BB}|$ the system goes back to the phase without compensation.
This is in agreement with the discussion presented above for Fig. \ref{fig:TvsJAB}
and it is not seen on the behavior of the bilayer
for either AP \cite{balcerzak2014ferrimagnetism} or MC \cite{diaz2016monte}.

It is possible to compare Fig. \ref{fig:phase} with
the phase diagram for the same system and same parameters obtained with the pair approximation,
presented in Fig. 2(a) of Ref. \onlinecite{szalowski2014normal}.
We first notice that the area occupied by the ferrimagnetic phase with compensation (to the left of the curve)
is consistently greater for MC than for the AP.
The lines also become more vertical as $p$ increases in the MC approach
and, although the reentrant behavior discussed in the last paragraph is also
present in the PA, it is only visible at $p=0.5$ in Fig. 2(a) of Ref. \onlinecite{szalowski2014normal},
while it is clearly present in MC for all values of $p$ presented in Fig. \ref{fig:phase}.
The PA results also indicate that there is no compensation whatsoever
for $p=0.6$ and $J_{AB}/J_{BB}$ below $\approx -0.86$
and for $p=0.5$ and $J_{AB}/J_{BB}$ below $\approx -0.36$,
whereas the MC results show we always have a phase with compensation
for $-1.0\leq J_{AB}/J_{BB}<0.0$ and the concentrations considered in Fig. \ref{fig:phase}.

\section{Conclusion}\label{conclusion}

In this work we have applied Monte Carlo simulations to study
the magnetic behavior of an Ising multilayer composed of alternated non-equivalent planes
of two types, \textbf{A} and \textbf{B}.
Both \textbf{A} and \textbf{B} intralayer couplings are ferromagnetic while
the interlayer couplings are antiferromagnetic. 
Additionally, only the \textbf{B} layers are site-diluted.
The simulations were performed using the Wolff algorithm
and the data were analyzed with multiple histogram reweighting and finite-size scaling methods.
Our main goal is to obtain the conditions of existence of a compensation point for the system,
i. e., a temperature $T_{comp}$ where the total magnetization is zero below the critical point $T_c$.

We studied simple cubic lattices with linear sizes $L\leq 100$
to allow for a precise evaluation of both $T_c$ and $T_{comp}$
for a wide range of values of each of the Hamiltonian parameters.
The results for the multilayer are compared with those for the bilayer
reported in both Monte Carlo \cite{diaz2016monte} and pair approximation \cite{balcerzak2014ferrimagnetism} approaches.
We see that the compensation phenomenon in the multilayer is favored by small but non-null dilutions
and by large intralayer coupling asymmetry, as it is also the case for the bilayer.
Similarly, the effect is favored by a weak interlayer coupling, although the sensitivity
to this parameter is more pronounced for the 2d system than it is for the multilayer.
In addition, we notice that the behavior of the multilayer gets closer to the bilayer
as the interlayer coupling gets weaker. This agrees with the crossover that
happens at that limit, where the multilayer becomes a set of non-interacting two-dimensional systems.

A summary of the results is then depicted in a convenient way on $J_{AB}\times J_{AA}$
diagrams for several values of site concentration.
These diagrams are compared with the PA results from Ref. \cite{szalowski2014normal} for the same model
and we emphasize that the MC and PA results
are considerably different, both quantitatively and qualitatively.

Work is now underway to accurately determine the critical exponents of the model,
as well as to extend the analysis to consider continuous spin symmetry, i. e., Heisenberg spins.

\begin{acknowledgments}
This work has been partially supported by Brazilian Agency CNPq.
\end{acknowledgments}

%

\newpage

\end{document}